\documentclass[
reprint,
amsmath,
prl,
superscriptaddress,
noeprint,
]{revtex4-1}
\usepackage{siunitx}
\usepackage{graphicx}
\usepackage{dcolumn}
\usepackage{hyperref}
\usepackage{braket}
\usepackage{xfrac}
\usepackage[usenames,dvipsnames]{color}
\usepackage{tikz-cd}
\usepackage{float}
\usepackage[nameinlink,capitalise]{cleveref}
\usepackage{mathtools}
\usepackage{amsmath}
\DeclareSIUnit\gauss{G}
\DeclareSIUnit{\au}{{a.u.}}
\hypersetup{
colorlinks=true,
linkcolor=blue,
filecolor=magenta,
urlcolor=cyan,
}
\begin{document}
\title{Phase-locking between different partial-waves in atom-ion spin-exchange collisions
}

\author{Tomas Sikorsky}
\affiliation{Department of Physics of Complex Systems, Weizmann Institute of Science, Rehovot 7610001, Israel.}
\email{tomas.sikorsky@weizmann.ac.il}
\author{Masato Morita}
\affiliation{Department of Physics, University of Nevada, Reno, NV, 89557, USA}
\author{Ziv Meir}
\affiliation{Department of Physics of Complex Systems, Weizmann Institute of Science, Rehovot 7610001, Israel.}
\author{Alexei A. Buchachenko}
\affiliation{Skolkovo Institute of Science and Technology, Skolkovo Innovation Center, Building 3, Moscow 143026, Russia}
\affiliation{Institute of Problems of Chemical Physics RAS, Chernogolovka, Moscow Region 142432, Russia}
\author{Ruti Ben-shlomi}
\affiliation{Department of Physics of Complex Systems, Weizmann Institute of Science, Rehovot 7610001, Israel.}
\author{Nitzan Akerman}
\author{Edvardas Narevicius}
\affiliation{Department of Chemical Physics, Weizmann Institute of Science, Rehovot 7610001, Israel.}
\author{Timur V. Tscherbul}
\affiliation{Department of Physics, University of Nevada, Reno, NV, 89557, USA}
\author{Roee Ozeri}
\affiliation{Department of Physics of Complex Systems, Weizmann Institute of Science, Rehovot 7610001, Israel.}

\date{\today}
\begin{abstract}
We present a joint experimental and theoretical study of spin dynamics of a single $^{88}$Sr$^+$ ion colliding with an ultracold cloud of Rb atoms in various hyperfine states. While spin-exchange between the two species occurs after 9.1(6) Langevin collisions on average, spin-relaxation of the Sr$^+$ ion Zeeman qubit occurs after 48(7) Langevin collisions which is significantly slower than in previously studied systems due to a small second-order spin-orbit coupling. Furthermore, a reduction of the endothermic spin-exchange rate was observed as the magnetic field was increased. Interestingly, we found that, while the phases acquired when colliding on the spin singlet and triplet potentials vary largely between different partial waves, the singlet-triplet phase difference, which determines the spin-exchange cross-section, remains locked to a single value over a wide range of partial-waves which leads to quantum interference effects.
\end{abstract}

\maketitle
In recent years, the research of atom-ion collisions has entered the ultracold regime. Since the first demonstrations~\cite{Kohl2009, Denschlag2010}, laser-cooled atom-ion hybrid systems have matured into a successful field of research~\cite{willitsch2014ion}, and many new phenomena have been observed. Examples include, state and spin controlled charge-exchange reactions~\cite{Ratschbacher2012a, Sikorsky2018}, molecule formation~\cite{hall2011light,wolf2017state}, the emergence of power-law energy distributions and non-equilibrium dynamics~\cite{Meir2016c, meir2018direct}, bifurcation of ion energies~\cite{Schowalter2016}, sympathetic and swap cooling~\cite{meir2018direct,Ravi2012a}, three-body reactions~\cite{Harter2012}, as well as spin exchange and spin relaxation in atom-ion collisions~\cite{Ratschbacher2013,Sikorsky2018,furst2017dynamics}.

Spin dynamics in atom-ion systems is particularly interesting. A single spin-1/2 ion immersed in a cloud of ultracold spin-polarized atoms realizes the model of a controllable qubit coupled to a well-defined and adjustable environment~\cite{Leggett1987,Prokofev2000}. Such a system can, for example, be used to model spin-impurities in the solid-state~\cite{Bissbort2013}. Understanding and controlling atom-ion spin dynamics is also essential for applications such as generating atom-ion entanglement or quantum gates~\cite{Secker2016, Doerk2010}. Previous experiments which investigated the spin dynamics of a single ion interacting with an ultracold gas used Yb$^+$/Rb~\cite{Ratschbacher2013} Yb$^+$/Li~\cite{furst2017dynamics} and Sr$^+$/Rb mixtures~\cite{Sikorsky2018}.
% A single isolated spin-1/2 particle interacting with its surroundings was theoretically studied\cite{Leggett1987,Prokofev2000}.

Here, we present a joint experimental and theoretical study of the spin dynamics of a single Sr$^+$ Zeeman qubit immersed in a spin-polarized bath of Rb atoms at mK temperature. Similarly to previous experimentally used species, both $^{87}$Rb and $^{88}$Sr$^+$ have a single electron in the valence shell. We carried out our measurements with Rb prepared in different hyperfine spin states and at two different magnetic fields. We found that similarly to other systems, spin dynamics is governed by the competition between spin-exchange (SE) and spin-relaxation (SR) processes. However, here, SR was found to be slow compared with the Langevin collision rate (\SI{48\pm7}{\tau_L}). This makes the Sr$^+$/Rb mixture promising for studying spin-dependent interactions. We found quantitative agreement between experimental observation and quantum scattering coupled-channel calculations based on {\it ab initio} potentials. Furthermore, we found that while at mK temperature multiple partial-waves ($\sim$15) contribute to the SE cross-section, the phase-difference acquired between the spin singlet and triplet incoming channels is the same for all partial-waves involved.

This new effect, which we term partial-wave phase-locking, leads to a dramatic sensitivity of the SE cross-section to a variation of the singlet-triplet energy gap even in the multiple partial-wave regime. Similar sensitivity was also recently observed in~\cite{furst2017dynamics}. We show that the physical origin of partial-wave phase-locking is the short-range nature of the spin-exchange interaction, which makes it independent of the orbital angular momentum of the collision complex that dominates long-range dynamics. This implies that, analogously to near-resonant charge exchange~\cite{BoGao12}, SE collision dynamics over a wide range of collision energies can be completely characterized by three parameters~\cite{furst2017dynamics}: the singlet and triplet scattering lengths, and the atomic polarizability. Our results open up the possibility to efficiently control binary spin-exchange collisions in hot atomic and molecular gases by varying the singlet-triplet energy gap or the reduced mass of the collision complex. As an illustration, we predict a substantial (three-fold) isotope effect for SE Sr$^+$-Rb collisions in the multiple partial-wave regime, which should be easily observable in near-future experiments.

\begin{figure}[!t]
\includegraphics[width=\linewidth]{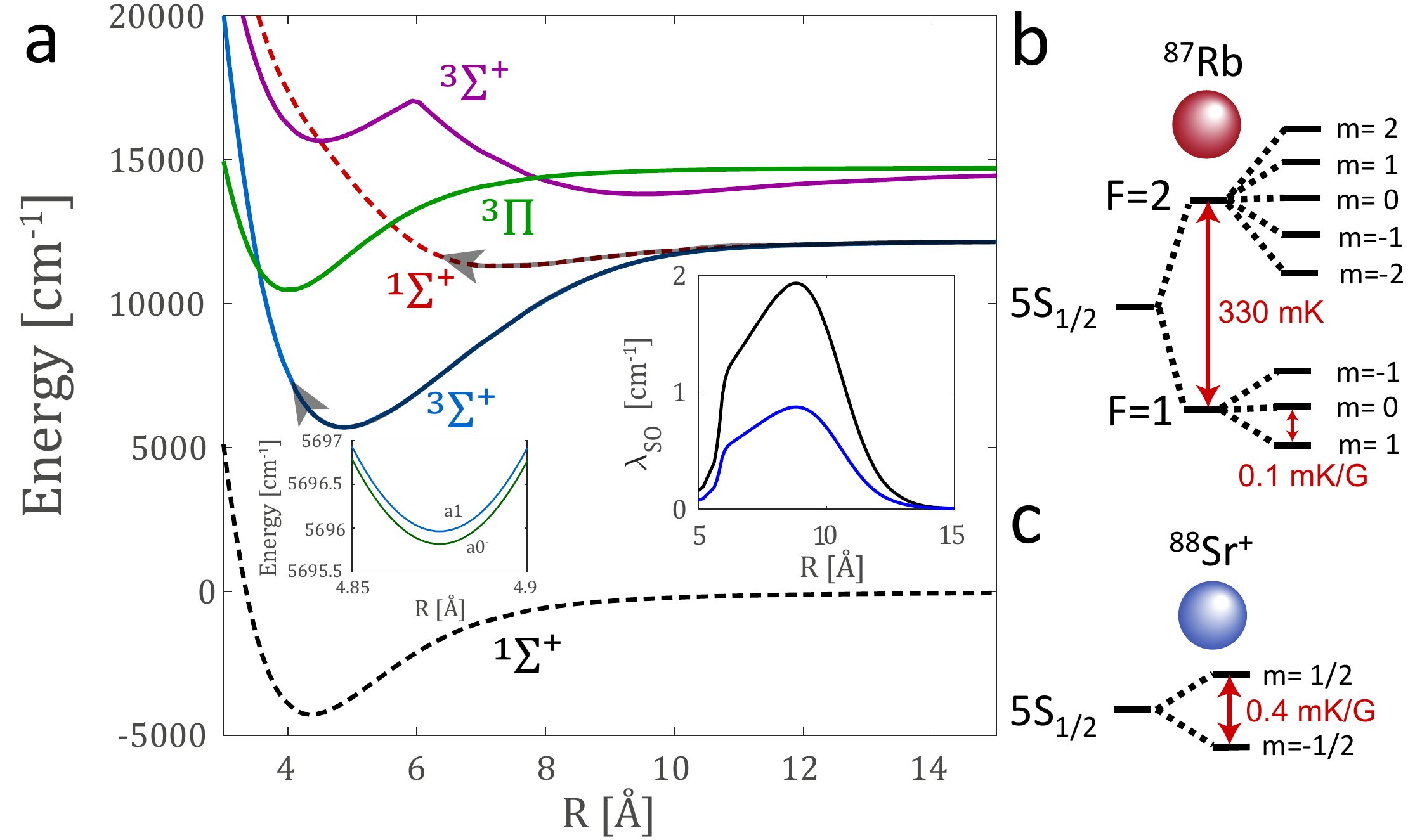}
\caption{\label{FIG. 1.}(a) Relevant PECs of the (RbSr)$^+$ complex~\cite{Aymar2011a}. Left inset shows the splitting of the $^{3}\Sigma^+$ state into a0$^-$ and a1 components. Right inset shows the $\lambda_{\mathrm{SO}}(R)$, the {\it ab initio} (black) and scaled (blue). (b-c) Level structure of the $^{88}$Sr$^+$ and the $^{87}$Rb ground states.}
\end{figure}

In our experiment~\cite{Meir2017} a single Sr$^+$ ion is trapped in a linear Paul trap, ground-state cooled to \SI{\sim40}{\mu\K} and spin-polarized to a state either parallel (m=1/2) $\ket{\uparrow}_{\mathrm{Sr}^+}$ or antiparallel (m=-1/2) $\ket{\downarrow}_{\mathrm{Sr}^+}$ to the magnetic field. Rb atoms are trapped in an optical trap, evaporatively cooled to \SI{\sim3}{\mu\K} and prepared in one of the spin states of the hyperfine manifold $\ket{F,m_{\mathrm{F}}}$ of the electronic ground state \cref{FIG. 1.}b. Interactions between the ion and the Rb cloud are initiated by moving the optical trap to overlap with the trapped-ion. After various interaction times, the atoms are released from the trap, and the ion spin projection along the quantization axis is measured using electron shelving on an optical clock transition, followed by state-selective fluorescence~\cite{SM,Keselman2011}.

To start our experimental investigation we polarize our ion and atoms to the $\ket{2,-2}_\mathrm{Rb}\otimes\ket{\downarrow}_{\mathrm{Sr}^+}$ ``stretched" state.
This state belongs to the triplet manifold only and is therefore fully protected against SE. Here, any spin dynamics we observe is solely due to SR processes.

\begin{figure}
\includegraphics[width=\linewidth]{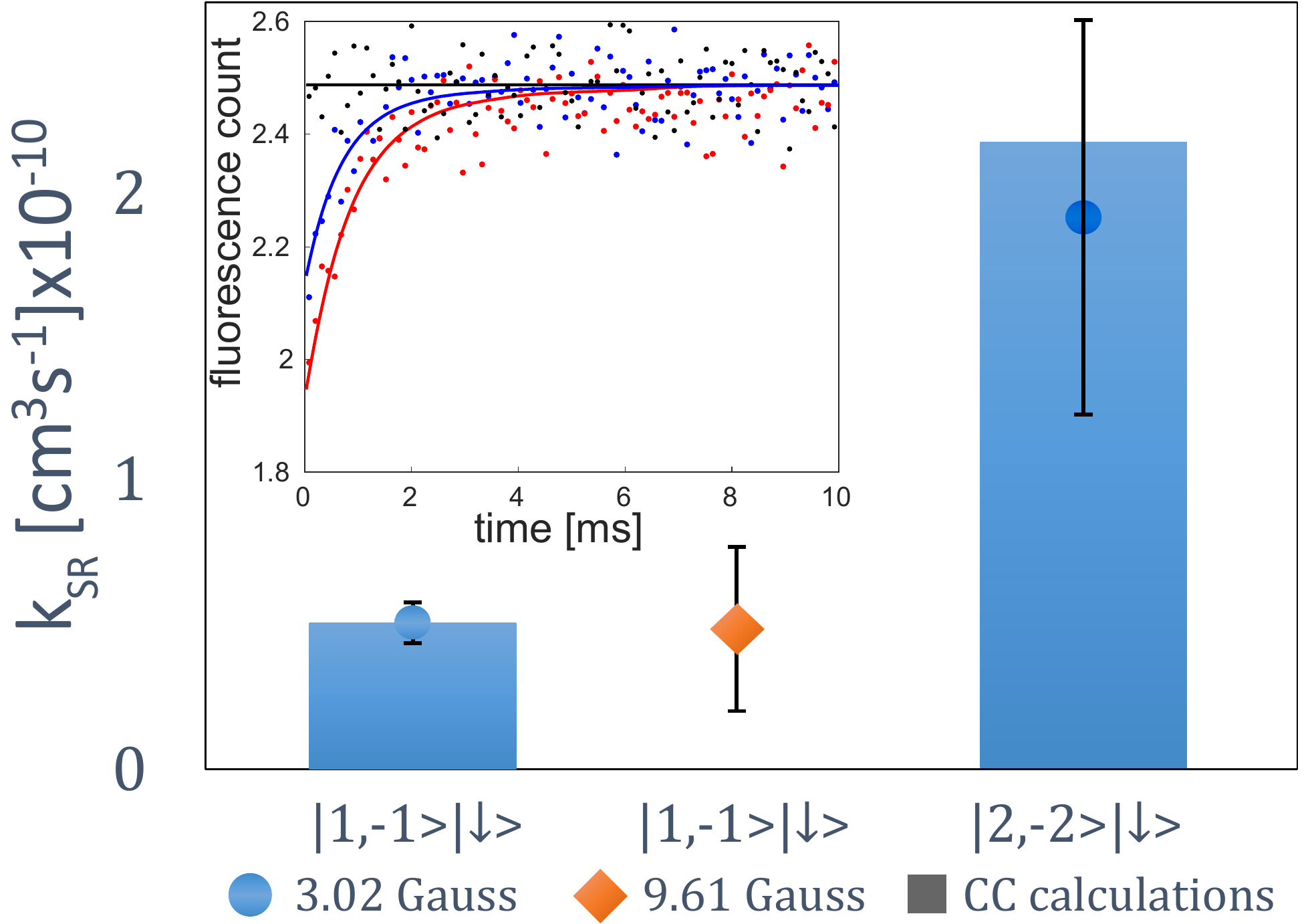}
\caption{\label{FIG. 4.} Experimental spin-relaxation rate constants ($k_{SR}$) for various initial states and magnetic fields: $\ket{1,-1}_\mathrm{Rb}\otimes\ket{\downarrow}_{\mathrm{Sr}^+}\rightarrow\ket{\mathrm{all}}_\mathrm{Rb}\otimes\ket{\uparrow}_{\mathrm{Sr}^+}\ \mathrm{and}\ \ket{2,-2}_\mathrm{Rb}\otimes\ket{\downarrow}_{\mathrm{Sr}^+}\xrightarrow{\text{SR}}\ket{F=1}_\mathrm{Rb}\otimes\ket{\mathrm{all}}_{\mathrm{Sr}^+}$. Bars represent the rate constants obtained from CC calculated cross-sections convolved with a corresponding energy distribution.
Inset shows Sr$^+$ ion fluorescence during Doppler cooling after \SI{500}{\milli\second} interaction time with $\ket{2,-2}$ (blue), $\ket{2,0}$ (red), and $\ket{1,-1}$ (black) Rb atoms. We extract the $\ket{2,-2}\ket{\downarrow}$ SR rate constant from a single parameter fit of the ion fluorescence curve, and the $\ket{1,-1}\ket{\downarrow}$ SR rate constant from the electron-shelving measurements (see \cref{FIG. 2.}).}
\end{figure}

We use ion thermometry to extract the SR rate of the system prepared in the $\ket{2,-2}_\mathrm{Rb}\otimes\ket{\downarrow}_{\mathrm{Sr}^+}$ state. Since most spin-relaxation channels from $\ket{2,-2}_\mathrm{Rb}\otimes\ket{\downarrow}_{\mathrm{Sr}^+}$ lead Rb atoms to the $F=1$ manifold, \SI{328}{\milli\kelvin} of energy is released in the process. This energy release heats the ion and leads to a higher steady-state temperature which we detect using Doppler cooling thermometry. From the fluorescence re-cooling curve (blue points in the inset of \cref{FIG. 4.}) we can extract the energy distribution of the ion at steady-state~\cite{Sikorsky2017}. Through a comparison with a molecular dynamics simulation the probability that hyperfine energy is released in a collision, $p_{HF}$, is extracted~\cite{SM}. 

Our data indicates $p_{HF} = 0.079\pm0.028$, which translates to a hyperfine energy-release rate of once every $13(5) \tau_L$, shown in \cref{FIG. 4.}. Furthermore, when initializing Rb in the $\ket{2,0}_\mathrm{Rb}$ state we get a higher steady-state temperature (red points in the inset of \cref{FIG. 4.}), consistent with SE processes adding to the hyperfine energy release rate.

\begin{figure}
\includegraphics[width=\linewidth]{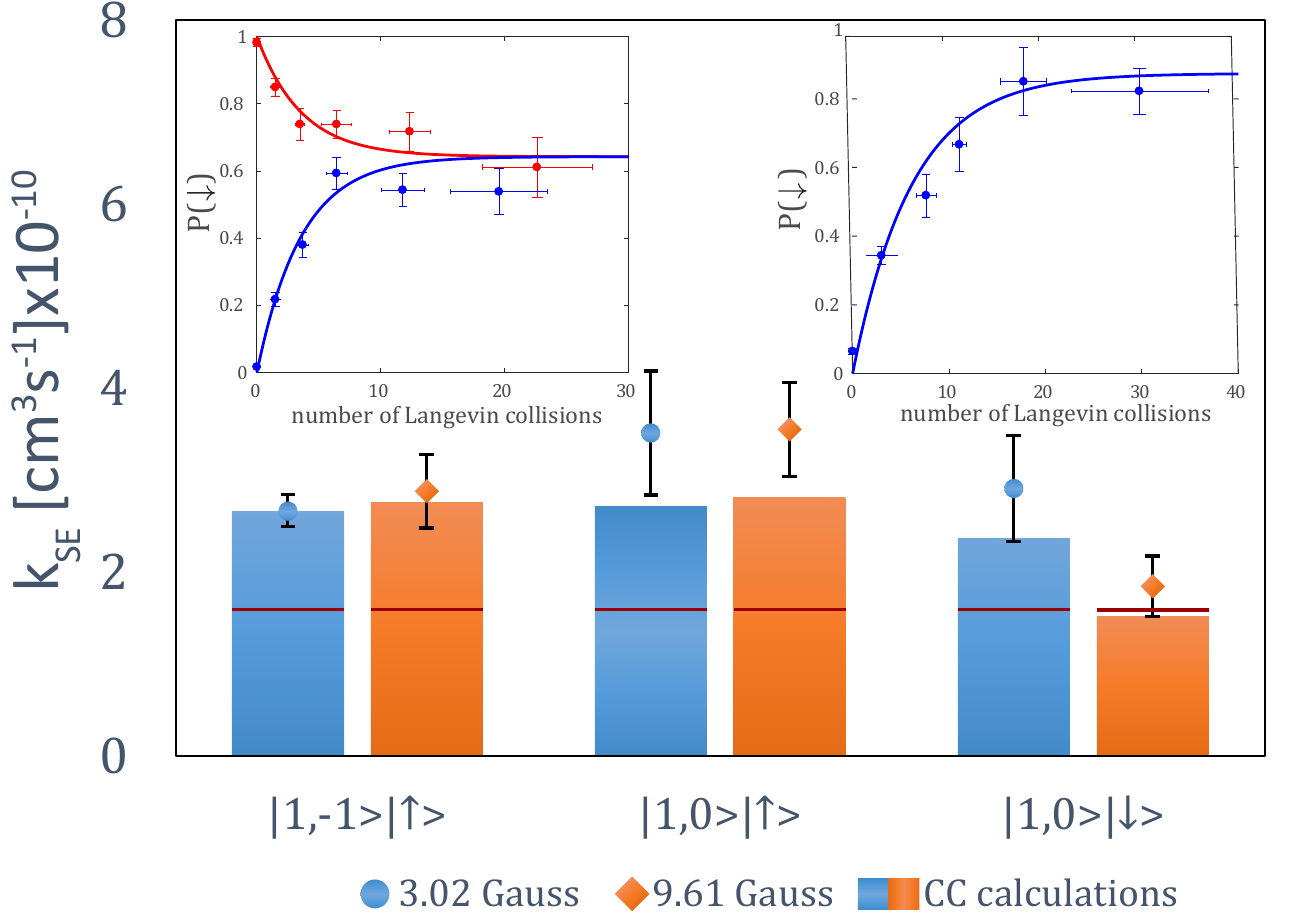}
\caption{\label{FIG. 2.}Experimental spin-exchange rate constants ($k_{SE}$) for various initial states and magnetic fields. Bars represent the rate constants obtained from CC calculated cross-sections convolved with a Tsallis energy distribution. Red line shows the RPA results, showing a clear disagreement with most of our measured rates. Insets show an example of the raw experimental data that were used to extract the rate constant. The left inset shows the evolution of the ion spin state prepared in either $\ket{\uparrow}_{\mathrm{Sr}^+}$ or $\ket{\downarrow}_{\mathrm{Sr}^+}$ state colliding with atoms in $\ket{1,0}_\mathrm{Rb}$ state. Right inset shows ion initialized in the $\ket{\uparrow}$ state colliding with atoms in $\ket{1,-1}_\mathrm{Rb}$. Both insets are for $B=\SI{9.61}{\gauss}$.}
\end{figure}

The increased ion temperature makes direct ion spin measurements using electron-shelving techniques difficult~\cite{SM}, as the ion is quickly heated out of the Lamb-Dicke regime. We, therefore, turn to measuring spin dynamics when Rb is initialized in the $F=1$ hyperfine ground manifold. We start by initializing Rb in the $\ket{1,-1}_\mathrm{Rb}$ state. Here we expect both SE and SR processes to play a role. When initializing Sr$^+$ in the $\ket{\uparrow}_{\mathrm{Sr}^+}$ state, SE flips its spin to $\ket{\downarrow}_{\mathrm{Sr}^+}$. However, when initializing in the $\ket{\downarrow}_{\mathrm{Sr}^+}$ state, SE transfers Rb to the $F=2$ hyperfine manifold. This process is energetically suppressed due to the $\SI{328}{\milli\kelvin}$ hyperfine energy gap. This asymmetry in SE collisions with Rb tends to polarize the ion spin to align with that of the atoms. An example of such spin dynamics is shown in the inset of \cref{FIG. 2.}. The steady-state polarization of the ion spin is determined by the ratio of SE to SR rates. We extracted both rates by comparing the measured Sr$^+$ ion spin dynamics to rate equations~\cite{SM}. 

In another measurement we prepared Rb atoms in the $\ket{1,0}$ state with ions in either $\ket{\uparrow}$ or $\ket{\downarrow}$. Here, SE can work both ways: $\ket{1,0}_\mathrm{Rb}\otimes\ket{\uparrow}_{\mathrm{Sr}^+}\rightarrow\ket{1,1}_\mathrm{Rb}\otimes\ket{\downarrow}_{\mathrm{Sr}^+}\ \mathrm{and}\ \ket{1,0}_\mathrm{Rb}\otimes\ket{\downarrow}_{\mathrm{Sr}^+}\rightarrow\ket{1,-1}_\mathrm{Rb}\otimes\ket{\uparrow}_{\mathrm{Sr}^+}$. The evolution of the ions spin for both initial states is shown in the inset of \cref{FIG. 2.}. As seen, the steady state polarization of the ion spin in this case is \SI{0.64\pm0.01}. The deviation from the expected value of 0.5 is due to small imbalance between the endo- and exo-thermic SE rates (see \cref{FIG. 2.}).

To theoretically explore SE collisions, we performed scattering calculations at various levels of sophistication, ranging from simple random-phase approximation (RPA), degenerate internal states approximation (DISA), to accurate coupled-channel (CC) calculations.

We begin with an expression of the SE cross-section in the DISA~\cite{Dalgarno1961,Dalgarno_Rudge_elasticapprox, DIS_PRA}, which assumes the degeneracy of internal states of Rb-Sr$^+$,
\begin{equation}
\label{dalgarno}
\small
\sigma^\mathrm{DISA}_{\mathrm{ex}}=\lvert\braket{\Psi_{\mathrm{i}}\rvert\mathrm{\mathbf{\hat{S}}^{(Rb)}\cdot\mathbf{\hat{S}}^{(Sr^{+})}}\rvert\Psi_{\mathrm{f}}}\rvert^2\cdot\frac{4\pi}{k^2}\sum_{l=0}^\infty(2l+1)\sin^2(\Delta \eta_{l}),
\end{equation}
where $\mathbf{\hat{S}}$ is the electron spin operator, $\Psi_{\mathrm{i/f}}$ describe the initial/final state of Sr$^+$/Rb mixture, 
$\Delta\eta_l$ is the difference of the singlet~(s) and triplet~(t) scattering phase shifts $\eta_l^{s,t}$, $k$ is the wave-number and $l$ is the orbital angular momentum. 
According to \cref{dalgarno}, SE between Rb and Sr$^+$ can be thought as an interference of the scattering wavefunctions on $V_{s,t}$: the singlet ($A^1\Sigma^+$) and triplet ($a^3\Sigma^+$) potential energy curves (PECs) in \cref{FIG. 1.}.

To obtain the scattering phase shifts, we carried out one-dimensional scattering calculations based on the calculated PECs~\cite{Aymar2011a}
merged with the long-range form $V_\text{as}(R)=-C_4/R^4-C_6/R^6$ using the $C_n$ coefficients~\cite{Cn}. 
%As described below, a scaled singlet potential is used in this letter unless otherwise stated~\cite{SM}. 

In the Langevin energy regime, \cref{dalgarno} can be further approximated $(\sin^2(\Delta \eta_{l})=\sfrac{1}{2})$ which gives the RPA,~\cite{Levine2005}
\begin{equation}
\label{rpa}
\sigma^\mathrm{RPA}_{\mathrm{ex}}=\lvert\braket{\Psi_{\mathrm{i}}\rvert\mathrm{\mathbf{\hat{S}}^{(Rb)}\cdot\mathbf{\hat{S}}^{(Sr^{+})}}\rvert\Psi_{\mathrm{f}}}\rvert^2\cdot 2\sigma_\mathrm{L},
\end{equation}
where $\sigma_\mathrm{L}=2\pi\sqrt{C_4/E}$ is the Langevin cross-section~\cite{LG}. Thus, the RPA cross-section is independent of the details of the PECs.

To solve the full ion-atom scattering problem including spin-orbit (SO) coupling, the hyperfine interaction, and an external magnetic field, we carry out CC calculations~\cite{SM}. The second-order spin-orbit coupling coefficient $\lambda_{\mathrm{SO}}$ determines the strength of the effective spin-spin interaction between the electron spins of valence electrons of Rb and Sr$^+$. It was identified as the main source of spin non-conserving processes in the Yb$^+$/Rb system~\cite{SM,Tscherbul2016,Ratschbacher2013}.
 
We now turn to compare our measured SE and SR rate to the results of the different calculations. The measured SR rate when initializing in $\ket{1,-1}_\mathrm{Rb}\otimes\ket{\downarrow}_{\mathrm{Sr}^+}$ of \SI{48\pm7}{\tau_L} is shown in \cref{FIG. 4.}. To match the SR rate predicted by CC calculations to the measured value we scaled $\lambda_{\mathrm{SO}}$ by a factor 0.45. Ability to reproduce SO splitting of an order of few wavenumbers within a factor of two reveals the high accuracy of the present {\it ab initio} calculations. The {\it ab-initio} calculated and scaled $\lambda_{\mathrm{SO}}$ are shown in the inset of \cref{FIG. 1.}. The vertical bars in \cref{FIG. 4.} show the calculated SR rate using the scaled $\lambda_{\mathrm{SO}}$ for both cases where Rb is initialized in the $\ket{1,-1}_\mathrm{Rb}$ and $\ket{2,-2}_\mathrm{Rb}$ states. As seen, in the latter case CC calculations reproduce the measured hyperfine-energy release rate without additional adjustment of $\lambda_{\mathrm{SO}}$ parameter.

\begin{figure}
\includegraphics[width=\linewidth]{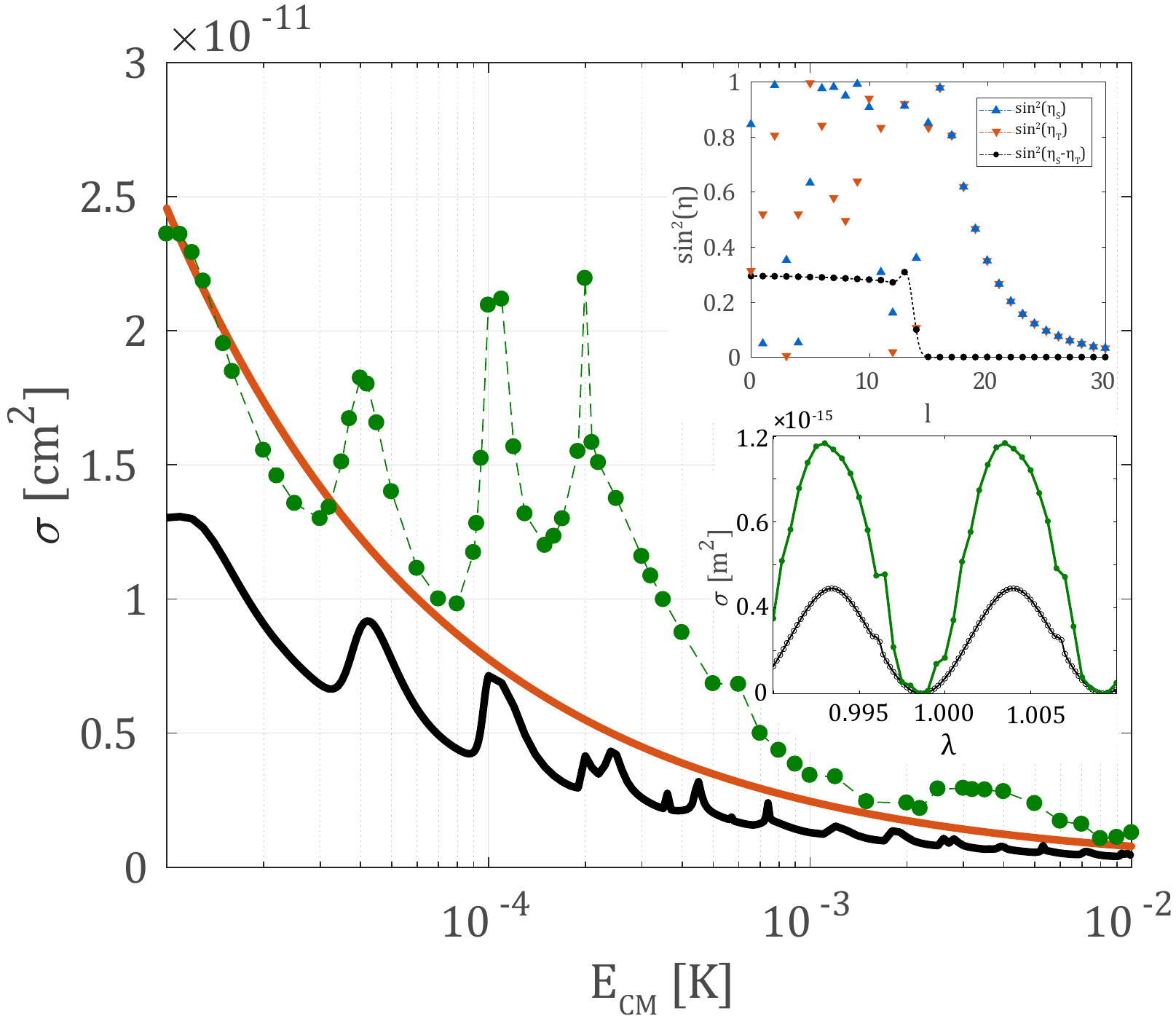}
\caption{\label{FIG. 3.}$\ket{1,-1}_\mathrm{Rb}\otimes\ket{\uparrow}_{\mathrm{Sr}^+}\rightarrow\ket{1,0}_\mathrm{Rb}\otimes\ket{\downarrow}_{\mathrm{Sr}^+}$ spin-exchange cross-section calculated at various levels of theoretical accuracy: RPA (red), DISA (green) and CC (green). The upper inset shows the difference of the scattering phase shifts, $\sin^2(\Delta\eta_l)$, as a function of $l$ obtained using the DISA at $E=\SI{1}{\milli\kelvin}$. The individual scattering phase shifts on the singlet and triplet PECs, $\sin^2(\eta_l^{s})$ and $\sin^2(\eta_l^{t})$, are shown by the blue and red filled-circles, respectively. The lower inset shows the spin-exchange cross-section at $E=\SI{1}{\milli\kelvin}$ as a function of the scaling parameter $\lambda$, where the black line is for DISA, and the green line is CC.}
\end{figure}

The SE rate we extract from this measurement corresponds to the $\ket{1,-1}_\mathrm{Rb}\otimes\ket{\uparrow}_{\mathrm{Sr}^+}\rightarrow\ket{1,0}_\mathrm{Rb}\otimes\ket{\downarrow}_{\mathrm{Sr}^+}$ transition. Here we observe SE once every \SI{9.1\pm0.59}{\tau_L} on average. To match the CC calculated value, obtained by convolving the CC cross-sections with Tsallis energy distribution, we tune the singlet-triplet gap by scaling the singlet potential~\cite{SM}. We find that the calculated SE cross-sections are highly sensitive to this scaling as shown in the inset of \cref{FIG. 3.}. We observe that the cross-sections oscillate periodically with full contrast, which suggests coherent partial-wave phase-locking as described below. A scaling factor of $\lambda=1.0005$ adequately matches our experimental result to CC theory.

We next analyze the case where Rb atoms are prepared in the $\ket{1,0}$ state. Because here SE and SR are experimentally indistinguishable, we assumed the same SR rates as for $\ket{1,-1}_\mathrm{Rb}$. The calculated SE rates agree with experimental rates~(\cref{FIG. 2.}), which justifies our initial parametrization of the Sr$^+$/Rb PEC.

As seen in \cref{FIG. 2.}, both theory and experiment indicate that for the increased magnetic field a difference between the endo- and exo- energetic SE cross-sections emerges. This is due to the Zeeman energetic barrier of $\SI{0.2}{\milli\kelvin\per\gauss}$, which increases with the magnetic field. Due to energy conservation increasing the magnetic field suppresses the SE cross-section at different energies~\cite{SM}. This demonstrates that magnetic field provides additional control over SE collisions.

To gain physical insight into the mechanism of SE collisions, we compare in \cref{FIG. 3.} the CC results at $B=\SI{3.02}{\gauss}$ with those obtained using the DISA and RPA. In all calculations the cross-sections scales as $\sigma\propto\sqrt{1/E}$ which confirms that spin-exchange is a Langevin process~\cite{Cote2000,Cote_Smith_SE_2003,LG}. We observe significant deviations of the results between CC and DISA, which comes from neglecting the hyperfine interaction in the DISA. In particular, the inter-channel coupling effect with the closed channels in the $F=2$ manifold of Rb is significant~\cite{SM}. On the other hand, the DISA cross-section correctly shows the positions of a series of shape resonances because those are determined by the shape of PECs rather than the inter-channel coupling. In the RPA, the resultant cross-section is systematically larger than the DISA cross-section which implies a dependence of the cross-section on the details of the PECs and a correlation between the difference of scattering phase shifts $\Delta\eta_l$ in different partial waves (\cref{dalgarno}). In what follows, we focus on these points. 

A remarkable result, which was also recently observed in~\cite{furst2017dynamics}, and is apparent in the lower inset of \cref{FIG. 3.} is the extreme sensitivity of SE cross-section to the potential scaling parameter $\lambda$ both with CC and DISA~\cite{SM}. This is unexpected, given that in the multiple partial wave regime one would expect random variations of $\sin^2(\Delta\eta_l)$ with $l$ in \cref{dalgarno} and the lack of sensitivity of scattering cross-sections to the PECs. To gain further insight, in \cref{FIG. 3.}, we plot the contributions sin$^2(\eta_l^{s,t})$ from the individual phase shifts on the singlet and triplet PECs along with their difference sin$^2(\Delta\eta_l)$. We observe that even though the $\sin^2(\eta_l^{s,t})$ for individual phase shifts change rapidly with $l$, their difference $\sin^2(\Delta\eta_l)$ remains constant and drops to zero after the height of the centrifugal barrier exceeds the collision energy $l>\tfrac{1}{\hbar}(16\mu^2C_4E)^{\sfrac{1}{4}}$. The phase-locking is responsible for the unexpected giant oscillations in SE cross-sections with respect to the scaling parameter $\lambda$.

\begin{figure}
\includegraphics[width=\linewidth]{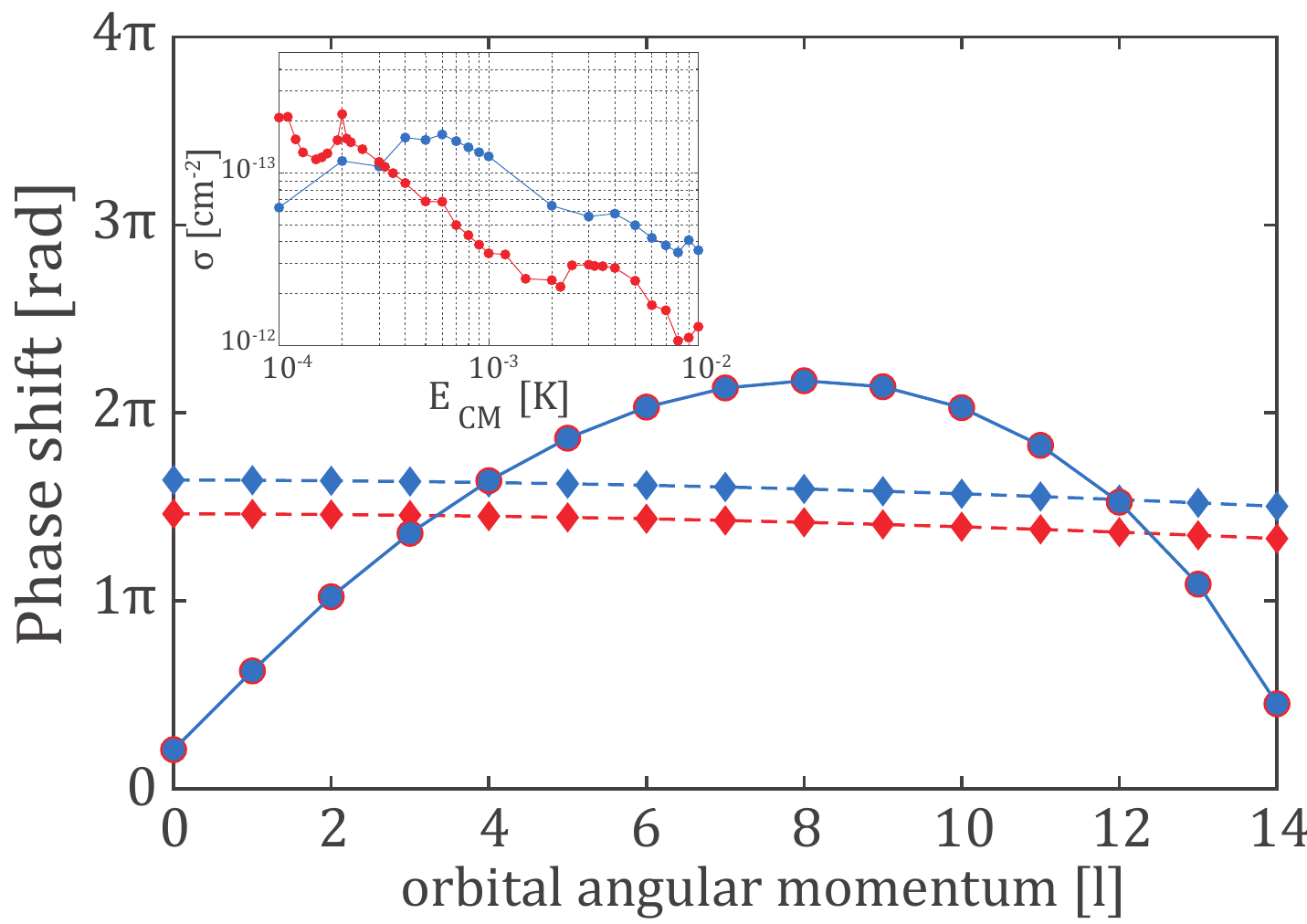}
\caption{\label{FIG. WKB.}
Short- (diamonds) and long- (circles) range phase shifts as a function of $l$ for the scattering wave functions on the singlet (red) and triplet (blue) potential energy curves at $E=\SI{1}{\milli\kelvin}$ with $R_\mathrm{mid}=\SI{30}{\bohr}$. Modulus is $2\pi$ for short range and $4\pi$ for long range. Inset: isotope dependence of the spin-exchange cross-section with $^{88}\text{Sr}^+$ (red) and $^{86}\text{Sr}^+$ (blue) using CC calculations ($B=\SI{3.02}{\gauss}$).}
\end{figure}

To elucidate the origin of partial-wave phase-locking, we evaluate the scattering phase shift using the WKB approximation,~\cite{WKB_Langer}

\begin{equation}
\begin{split}
\eta_l^{s,t} =&\int_{R_0}^\infty \sqrt{k^2-\frac{(l+1/2)^2}{R^2}-U_{s,t}(R)}\ \mathrm{d}R\\*\label{wkb_phase}-&\int_{R'_0}^\infty \sqrt{k^2-\frac{(l+1/2)^2}{R^2}}\ \mathrm{d}R,
\end{split}
\end{equation}
where $k^2=2\mu E$, $U_{s,t}(R)=2\mu V_{s,t}(R)$, $R_0$ and $R'_0$ are the classical turning points in the presence and in the absence of the potential. Separating the integration range in \cref{wkb_phase} into the short-range and long-range parts, we define the short-range phase shift $\eta^{s,t,SR}_l=\int_{R_0}^{R_\mathrm{mid}}\sqrt{...}\ \mathrm{d}R-\int_{R'_0}^{R_\mathrm{mid}}\sqrt{...}\ \mathrm{d}R$ and the long-range phase shift $\eta^{s,t,LR}_l=\int_{R_\mathrm{mid}}^\infty \sqrt{...}\ \mathrm{d}R- \int_{R_\mathrm{mid}}^{\infty}\sqrt{...}\ \mathrm{d}R$, with $R_\mathrm{mid}=\SI{30}{\bohr}$. Since $V_s(R) \simeq V_t(R)$ at long-range, $\eta^{s,LR}_l \simeq \eta^{t,LR}_l$ as illustrated in \cref{FIG. WKB.}. Thus, $\Delta\eta_l$ is determined entirely by the difference of the short-range phase shifts as $\Delta\eta_l \simeq \eta^{s,SR}_l -  \eta^{t,SR}_l$.

In \cref{FIG. WKB.}, we observe that the individual short-range phase shifts, as well as their difference, depend on $l$ only very weakly due to the small magnitude of the centrifugal potential compared with the potential well depth at short range. The physical origin of phase-locking can thus be attributed to the short-range nature of the SE interaction and to the large potential well-depth which renders short-range physics independent of $l$~\cite{BoGao01}. In other words, centrifugal forces play an important role only at atom-ion separations at which spin-exchange interaction is negligible. A sensitivity to the singlet-triplet gap could lead to a significant difference in the SE rate between different isotopes. The inset of \cref{FIG. WKB.} shows a comparison between the calculated SE cross-sections of $^{86}$Sr$^+$ and $^{88}$Sr$^+$ colliding with $^{87}$Rb atoms indeed predicting a three-fold ratio between the cross-sections over a wide range of energies.

In conclusion, we have studied experimentally and theoretically, the spin dynamics of a single $^{88}$Sr$^+$ ion immersed in a spin-polarized cloud of $^{87}$Rb atoms. We have shown that, for this mixture, spin dynamics is dominated by spin-exchange while spin relaxation is suppressed due to weak spin-orbit coupling. Our measurements are in excellent agreement with theoretical calculations. Furthermore, by varying the ambient magnetic field, we were able to control the rate of endothermic spin-exchange. This ability, together with slow spin relaxation, suggests that working at high magnetic fields it would be possible to freeze spin populations in this particular mixture for a long time. Interestingly, we found that at our collision energy, the collision cross-section is largely independent of the partial wave involved, leading to coherent oscillation in the spin-exchange rate as atomic potentials are varied. A future measurement of the spin-exchange rate using a different isotope of the Sr ion, $^{86}$Sr$^+$, would verify this phase-locking effect.

\begin{acknowledgments}
This work was supported by the Crown Photonics
Center, ICore-Israeli excellence center circle of light, the
Israeli Science Foundation, the U.S.-Israel Binational
Science Foundation, and the European Research Council (consolidator grant 616919-Ionology). The work at UNR was supported by NSF grant PHY-1607610.
\end{acknowledgments}
\clearpage
\begin{center}
	\textbf{\Large Supplemental Material}
\end{center}
\begin{center}
	\textbf{Experimental procedure}
\end{center}

A more detailed description of the experimental apparatus can be found in a recent publication~\cite{Meir2017}. We initialize $^{87}$Rb atoms in the F=1 state of the hyperfine manifold and temperature T\SI{\approx3}{\micro\kelvin} in an optical lattice (\SI{1064}{\nano\meter} YAG laser). We transfer the atoms over \SI{25}{\centi\meter} to the ion's chamber where they are loaded into a crossed dipole trap ($[\omega_{x},\omega_{y},\omega_{z}]$=2$\pi\times$[0.61, 0.6, 0.1] kHz) \SI{50}{\micro\meter} above the Sr$^+$ ion. Here, $\sim$10$^5$ atoms are spin-polarized using a combination of resonant microwave pulses and \SI{780}{\nano\meter} laser light. The polarization fidelity is above $>$$99\%$.
The Sr$^+$ ion is trapped with a rf linear segmented Paul trap with secular trap frequencies of $\omega=2\pi\times$[0.8, 1, 0.4] MHz for the two radial and the axial mode respectively. We perform ground state cooling and spin state preparation using a narrow linewidth \SI{674}{\nano\meter} laser on the $S_{1/2}\rightarrow D_{5/2}$ quadrupole transition.
To overlap the atoms with the ion, we move the crossed dipole trap onto the ion position.

During atom-ion interaction all lasers beams are mechanically blocked except for the off-resonant dipole-trap lasers at \SI{1064}{\nano\meter}. After the desired interaction time, we shut off the dipole-trap lasers which results in a free-fall expansion of the atoms. At the end of the atoms time-of-flight, we detect their number and temperature using the absorption-imaging technique. The measured density and temperature are used for the atom density estimation. We then perform Rabi carrier spectroscopy on the narrow $S_{1/2}\rightarrow D_{5/2}$ optical quadrupole transition~\cite{Meir2017} and Doppler cooling thermometry~\cite{Sikorsky2017} on the dipole $S_{1/2}\rightarrow P_{1/2} \leftarrow D_{3/2}$ closed-cycle transitions.

\begin{center}
	\textbf{Rate equations}
\end{center}
The spin-exchange and spin-relaxation dynamics of the Sr$^+$ ion in the atomic bath is governed by two-level rate equations:
\begin{equation}
\label{rate1}
\dot{p_{\downarrow}}=\gamma_{SE}\cdot p_{\uparrow}+\gamma_{SR}\cdot (p_{\uparrow}-p_{\downarrow}),
\end{equation}
\begin{equation}
\begin{split}
\label{rate2}
\dot{p_{\uparrow}}=(\gamma_{SE}+\gamma_{SR})\cdot (p_{\downarrow}-p_{\uparrow}),\\
\dot{p_{\downarrow}}=(\gamma_{SE}+\gamma_{SR})\cdot (p_{\uparrow}-p_{\downarrow}).
\end{split}
\end{equation}
Here, \cref{rate1} is for atoms in $\ket{1,-1}_\mathrm{Rb}$ state and \cref{rate2} is for atoms in $\ket{1,0}_\mathrm{Rb}$ state. $\gamma_{SE}$ ($\gamma_{SR}$) are spin-exchange (spin-relaxation) constants and $p_{\uparrow}+p_{\downarrow}=1$. The collisional rate constant is defined as $k=1-e^{-\gamma}$.
\begin{center}
	\textbf{Tsallis energy distribution}
\end{center}
Due to micromotion-induced collisional heating~\cite{Cetina2012}, following $\sim$20 Langevin collisions, the ion develops a power-law energy distribution with most probable energy of $\SI{\sim1.7}{\milli\kelvin}$. The distribution was extracted from a molecular dynamics simulation~\cite{Meir2016c}. Despite the fact that both species are trapped, we theoretically treat the collision as occurring between two free particles. In this case, the energy in the center-of-mass frame is dominated by the ion energy. The energy of the atoms is significantly lower ($\SI{\sim3}{\micro\kelvin}$).
\begin{center}
	\textbf{Simulations}
\end{center}
We performed a hard-sphere-type collision simulation as described in~\cite{Meir2016c,Zipkes2011}. This simulation does not include the collision-induced micromotion due to lack of polarization $-\sfrac{1}{r^4}$ attractive force. In the experiment, the excess micromotion was compensated below $\SI{100}{\micro\K}$. To mimic the effect of collision-induced micromotion, we added the excess micromotion to obtain the same energy distribution as can be obtained with much more demanding simulation with polarization potential~\cite{Meir2016c,Meir2017} (see the blue curve on Supplementary figure 1).
\begin{figure}
	\includegraphics[width=\linewidth]{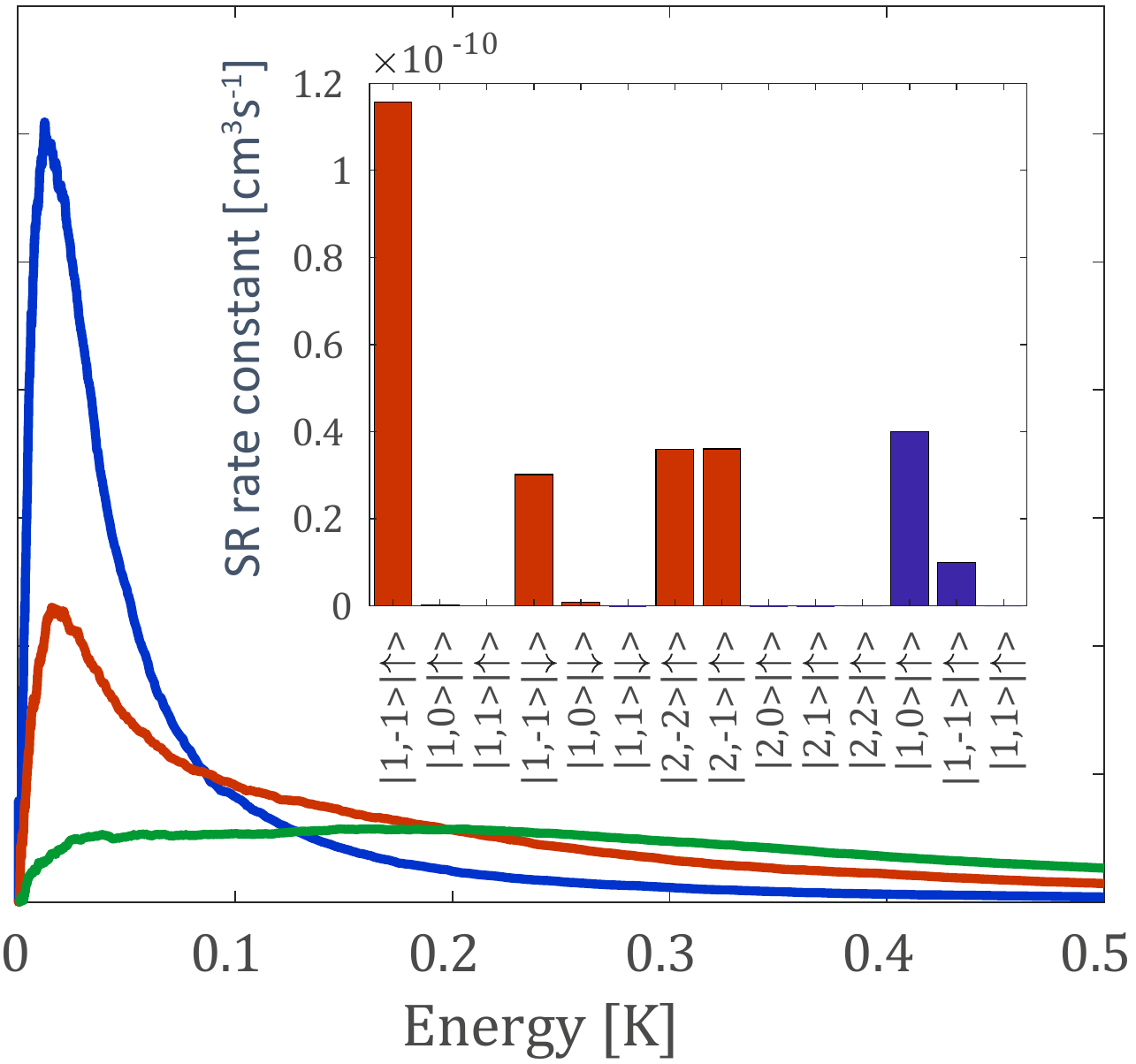}
	\caption{\label{Supplementary figure 1.}Supplementary figure 1. Molecular dynamics simulation of the ion’s energy distribution for different hyperfine energy release probabilities. Blue $p_{HF}=0\%$, red $p_{HF}=7.92\%$, green $p_{HF}=24\%$. The inset shows SR cross-section for atoms and ion initialized in $\ket{1,-1}_\mathrm{Rb}\otimes\ket{\downarrow}_{\mathrm{Sr}^+}$ (blue bars) and $\ket{2,-2}_\mathrm{Rb}\otimes\ket{\downarrow}_{\mathrm{Sr}^+}$ (red bars).}
\end{figure}
To introduce the effect of hyperfine energy release, during every hard-sphere collision there is a $p_H$ probability of increasing the ion energy by $\frac{\mu}{m_{\mathrm{Sr^+}}}\SI{328}{\micro\K}$. The steady-state distribution is obtained after 100 consecutive hard-sphere collisions by repeating this sequence 10000 times. Examples of the energy distributions for various $p_H$ are shown in Supplementary figure 1.

\begin{center}
	\textbf{Numerical solution of CC equations}
\end{center}
We carry out CC calculations using the following Hamiltonian for the collision complex~\cite{Tscherbul2016,JuliennePRA09,JulienneNJP11} as 
\begin{equation}
\hat{\mathcal{H}} = - \frac{1}{2\mu R}  \frac{\partial^2}{\partial R^2} R
+ \frac{\boldsymbol{\hat{l}}^2}{2\mu R^2} \allowbreak
+ \hat{\mathcal{H}}_{\mathrm{Rb}}
+ \hat{\mathcal{H}}_{\mathrm{Sr^+}}
+ \hat{\mathcal{H}}_{\mathrm{int}},
\label{eq:Heff}
\end{equation}
where $\mu$ is the reduced mass of the collision complex. The asymptotic Hamiltonian $\hat{\mathcal{H}}_i=\gamma_{i} \hat{\boldsymbol{\mathrm{I}}}_{i} \cdot \hat{\boldsymbol{\mathrm{S}}}_{i}+2 \mu_\mathrm{B} \mathbf{B} \cdot \hat{\boldsymbol{\mathrm{S}}}_{i}$ ($i=\mathrm{Rb},\mathrm{Sr^+}$) describes the hyperfine coupling
%electron and nuclear spins of the collision partners $\hat{\boldsymbol{\mathrm{S}}}_{i}$ and $\hat{\boldsymbol{\mathrm{I}}}_{i}$
and the interactions with an external magnetic field $\boldsymbol{\mathrm{B}}$, which are neglected in the DISA and RPA, where $\gamma_i$ is the hyperfine constant of the $i^\mathrm{th}$ atom, and $\mu_\mathrm{B}$ is the Bohr magneton. The ion-atom interaction operator $\hat{\mathcal{H}}_{\mathrm{int}}$ includes the electrostatic interaction $\hat{V}(R)$ (see \cref{FIG. 1.}), the magnetic dipole-dipole interaction $\hat{V}_\text{dd}(R)$, and the second-order spin-orbit interaction $\hat{V}_\text{SO}(R)$. 
%$\hat{V}(R)$ is given by $\hat{V}(R)=\sum^{}_{S, M_S} |S M_S \rangle V_S(R) \langle S M_S|$, where $M_S$ is the projection of electron spin $S$ of the collision complex on the $B$-field axis. 
%For the singlet $V_s(R)$, we use a scaled potential energy curve 
%rather than the original ab inito one 
%\cite{Aymar2011a}. 
%The value of the scaling parameter, $\lambda$, is determined such that CC reproduces an experimental rate constant for the $\ket{1,-1}_\mathrm{Rb}\otimes\ket{\uparrow}_{\mathrm{Sr}^+}\rightarrow\ket{1,0}_\mathrm{Rb}\otimes\ket{\downarrow}_{\mathrm{Sr}^+}$ process at $B=\SI{3.02}{\gauss}$
%(see Supplemental Material~\cite{SM}).
Spin relaxation is caused by the latter two terms~\cite{Tscherbul2016,SO_Hamiltonian},
\begin{equation}
\small
\begin{split}
&\hat{V}_\text{dd}(R)+\hat{V}_\text{SO}(R) \\ = &\sqrt{\frac{24\pi}{5}} \left[-\frac{\alpha ^2}{R^3}+\lambda_\mathrm{SO}(R)\right]  \sum^{}_{q} (-1)^q Y_{2,-q}^*(\hat{\bm{R}}) [\hat{\mathbf{S}}_\mathrm{Rb}\otimes \hat{\mathbf{S}}_\mathrm{Sr^+}]_{q}^{(2)},
\label{eq:Hdip}
\end{split}
\end{equation}
where $\alpha$ is the fine-structure constant and $[\hat{\mathbf{S}}_\mathrm{Rb}\otimes \hat{\mathbf{S}}_\mathrm{Sr^+}]_{q}^{(2)}$ is a spherical tensor product of $\hat{\mathbf{S}}_\mathrm{Rb}$ and $\hat{\mathbf{S}}_\mathrm{Sr^+}$. The second-order SO interaction is parametrized with $\lambda_\mathrm{SO}(R)$ (see below).

To solve the scattering problem defined with the Hamiltonian in \cref{eq:Heff}, we expand the wavefunction of the collision complex in a set of basis functions $\phi_n =|F\, m_F\rangle_\mathrm{Rb}\ |F\, m_F\rangle_{\mathrm{Sr}^+}\ |l\, m_{l}\rangle$,
which leads to a set of coupled-channel (CC) equations for the radial expansion coefficients $\mathcal{F}_{n}(R)$~\cite {JuliennePRA09,JulienneNJP11,Tscherbul2016}
\begin{equation}
\begin{split}
[ \frac{d^2}{dR^2} & -\frac{l(l+1)}{R^2} +2\mu E ]\mathcal{F}_n(R)  \\ & =
2\mu \sum^{}_{n'} \langle \phi_n | \hat{\mathcal{H}}_{\mathrm{a}} + \hat{\mathcal{H}}_{\mathrm{b}} + \hat{\mathcal{H}}_{\mathrm{int}} |\phi_{n'}\rangle\mathcal{F}_{n'}(R),
\label{eq:CC}
\end{split}
\end{equation}
where $|F\, m_F\rangle$ are the hyperfine states, $|l\, m_{l}\rangle$ are the eigenstates of $\boldsymbol{\hat{l}}^2$ and $\hat{l}_z$, and $E$ is the total energy.
Since the total angular momentum projection, $M = m_{F_{\mathrm{a}}}+m_{F_{\mathrm{b}}}+m_L$, is conserved, we integrate the CC equation independently for each value of $M$ from $R=\SI{5.0}{\bohr}$ to $10^4$ \SI{}{\bohr} in steps of \SI{0.002}{\bohr} using the modified log-derivative propagator method~\cite{prop}. Matching the solutions of \cref{eq:CC} to scattering boundary conditions yields the $S$-matrix, from which we obtain the inelastic cross-sections~\cite{JuliennePRA09, Tscherbul2016}. We employ CC basis sets containing 20-50 partial waves to ensure numerical convergence of the inelastic cross-sections over the range of collision energies $\SIrange[range-phrase = -]{1}{100}{\milli\kelvin}$. 

\begin{center}
	\textbf{Potential scaling}
\end{center}
Cross-sections of the Hamiltonian (\cref{eq:Heff}) can be obtained from CC calculations. However, the interaction part in the Hamiltonian has uncertainty in particular for systems containing heavy atoms. The uncertainty for the {\it ab initio} interaction potential is usually estimated around or larger than 5\%~\cite{Tomza_Verror, Wallis_Verror}. Thus, the scaling or re-fitting of calculated potentials have been performed to reproduce experimental properties, to follow physical assumptions, and to investigate the robustness of the results against the change of the potentials~\cite{Wallis_Verror, Hutson_Vscale, Brue_Vscale,furst2017dynamics,furst2017dynamics}.

In this letter, the calculated singlet potential energy curve $V_s(R)$~\cite{Aymar2011a} is scaled, such that the CC result match our experimental result for the spin-exchange rate for the $\ket{1,-1}_\mathrm{Rb}\otimes\ket{\uparrow}_{\mathrm{Sr}^+}\rightarrow\ket{1,0}_\mathrm{Rb}\otimes\ket{\downarrow}_{\mathrm{Sr}^+}$ process at $B=\SI{3.02}{\gauss}$, using a constant scaling factor $\lambda$ as
\begin{equation}
V'_s(R)= V_t(R)  + \lambda \Delta V(R),
\label{eq:scalepot}
\end{equation}
where $V_t(R)$ is the potential energy curve for the triplet, and $\Delta V(R)$ is a function of $R$ defined as the difference between $V_s(R)$ and $V_t(R)$, namely $\Delta V(R)=V_s(R)-V_t(R)$. Thus, $\lambda=1$ results in $V'_s(R)=V_s(R)$. With any finite value of $\lambda$, the asymptotic form of the singlet potential energy curve, which is determined by $C_4$ and $C_6$, is not changed by this scaling. We employ $\lambda=1.0005$ as the value closest to one and reproducing the experimental rate constant within the error bar both with the Tsallis and Boltzmann distributions. 

For the calculation of spin relaxation, $\lambda_\mathrm{SO}(R)$ is scaled by multiplying a constant factor of 0.45 to reproduce the experimental rate for the
$\ket{1,-1}_\mathrm{Rb}\otimes\ket{\downarrow}_{\mathrm{Sr}^+}\rightarrow\ket{1,0}_\mathrm{Rb}\otimes\ket{\uparrow}_{\mathrm{Sr}^+}$ process at $B=\SI{3.02}{\gauss}$.

\begin{center}
	\textbf{Spin-orbit coupling}
\end{center}
The second-order spin-orbit interaction couples molecular states with different total spin $S$ at short range, leading to efficient spin relaxation in heavy ion-atom collisions~\cite{Tscherbul2016}.
The second-order SO coupling term proportional to $\lambda_\mathrm{SO}(R)$ in the \cref{eq:Hdip} is due to the SO interaction between the $a^3\Sigma^+$ electronic state and the electronic states of $^3\Pi$ symmetry (see \cref{FIG. 1.})~\cite{SO_Hamiltonian}. 
To evaluate $\lambda_\text{SO}(R)$ we use the same procedure as described in our previous work~\cite{Tscherbul2016}. In brief, the SO matrix elements between the $^3\Sigma^+$ and the excited $^3\Pi$ electronic states are obtained using the complete active space multiconfigurational self-consistent field method followed by state-interacting SO configuration interaction calculations~\cite{SOCI} as implemented in the MOLPRO package of {\it ab initio} programs~\cite{MOLPRO_brief}. The non-relativistic part of the electronic Hamiltonian is parameterized by the potentials~\cite{Aymar2011a} shifted in energy to match the experimental threshold energies~\cite{NIST_ASD}.

Diagonalization of the Hamiltonian matrix gives the potential energy curves $V_s(R)$ and $V_t(R)$ for SO-coupled $a0^-$ and $a1$ components of the $a^3\Sigma^+$ state, respectively,
\begin{equation}
\lambda_\mathrm{SO}(R) = -\frac{2}{3}(V_s-V_t). 
\end{equation}
To get an insight into the expected accuracy of the {\it ab initio} SO calculations, it is useful to recall that perturbatively
\begin{equation}
\lambda_\mathrm{SO}(R) \propto \frac{|\langle ^3\Sigma^+|\hat{H}_\mathrm{SO}|^3\Pi\rangle|^2}{V(^3\Pi)-V(^3\Sigma^+)}, 
\end{equation}
where the SO matrix element in the numerator is of order of \SI{100}{\per\centi\meter}, while the difference of potential energies in the denominator is of order of \SI{15000}{\per\centi\meter}. So desirable small parameter $\lambda_\mathrm{SO}$ comes out as the ratio of two big numbers subjected by errors in the non-relativistic wave functions and SO integrals. 
%and the splitting between the $a0^-$ and $a1$ SO states is determined directly by diagonalization of the Hamiltonian.
\color{black}

\begin{center}
	\textbf{Effect of magnetic field}
\end{center}
Since DISA does not include the effect of an external magnetic field, we performed CC calculations for different magnetic fields. We found that the magnetic field affects the cross-section around the resonance peaks. Comparing CC at $B\rightarrow\SI{0}{\gauss}$ with DISA reveals that the missing of the interchannel coupling with the closed channels due to the hyperfine interaction is responsible for significant underestimation of the cross-section by DISA.
\begin{figure}
	\includegraphics[width=\linewidth]{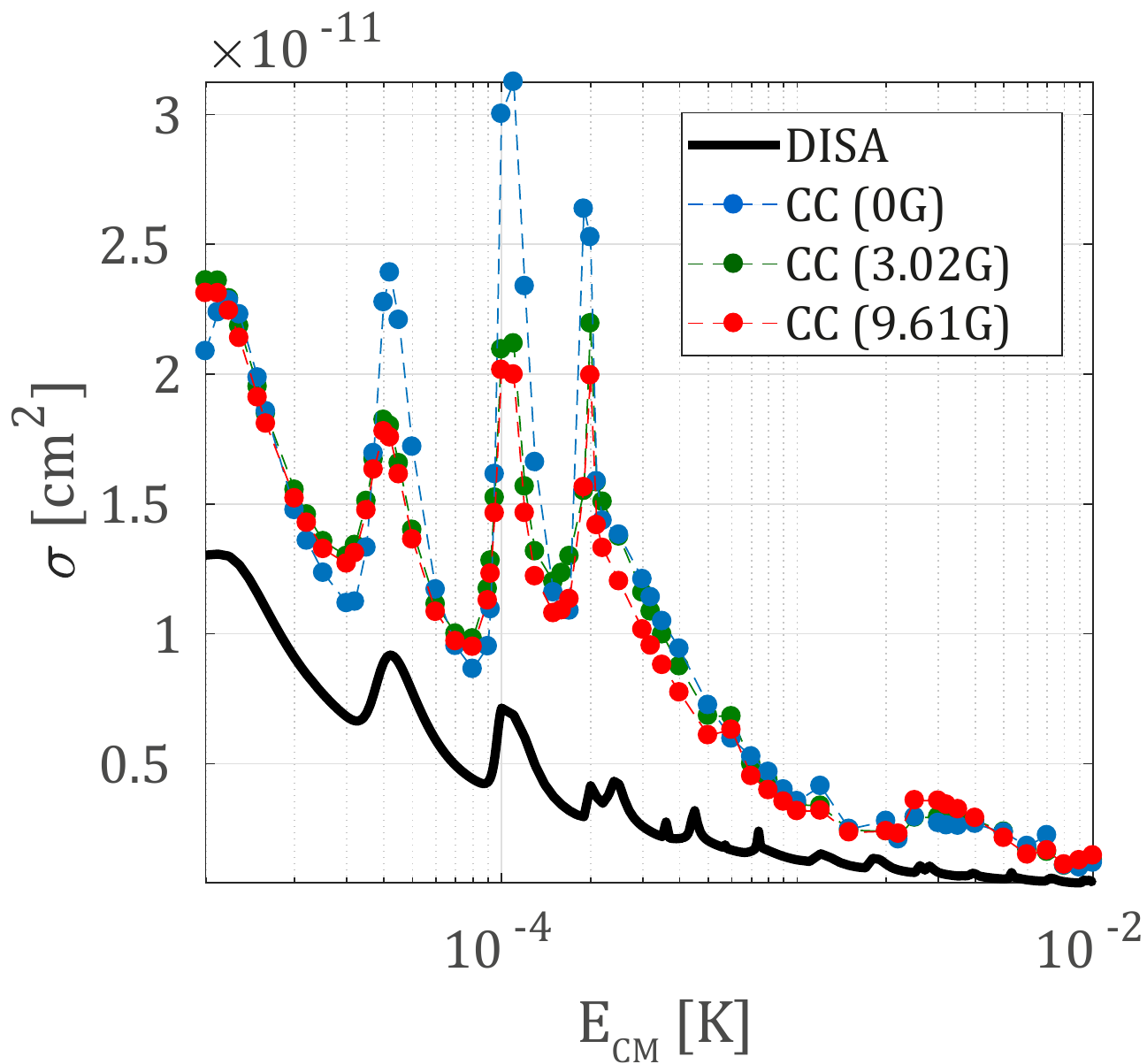}
	\caption{\label{Supplementary figure 2.}Supplementary figure 3. $\ket{1,-1}_\mathrm{Rb}\otimes\ket{\uparrow}_{\mathrm{Sr}^+}\rightarrow\ket{1,0}_\mathrm{Rb}\otimes\ket{\downarrow}_{\mathrm{Sr}^+}$ SE cross-section calculated with coupled-channel (CC) calculation at various magnetic fields. Black line compares these calculations with DISA.}
\end{figure}
While magnetic field does not affect the exoenergetic spin-exchange channels dramatically, it affects endoenergetic channels. The cross-section significantly drops for collision energies below the Zeeman energetic barrier, $E_{Zeeman}=\SI{0.2}{\milli\kelvin\per\gauss}$. This effect is simply due to the energy conservation.
\begin{figure}
	\includegraphics[width=\linewidth]{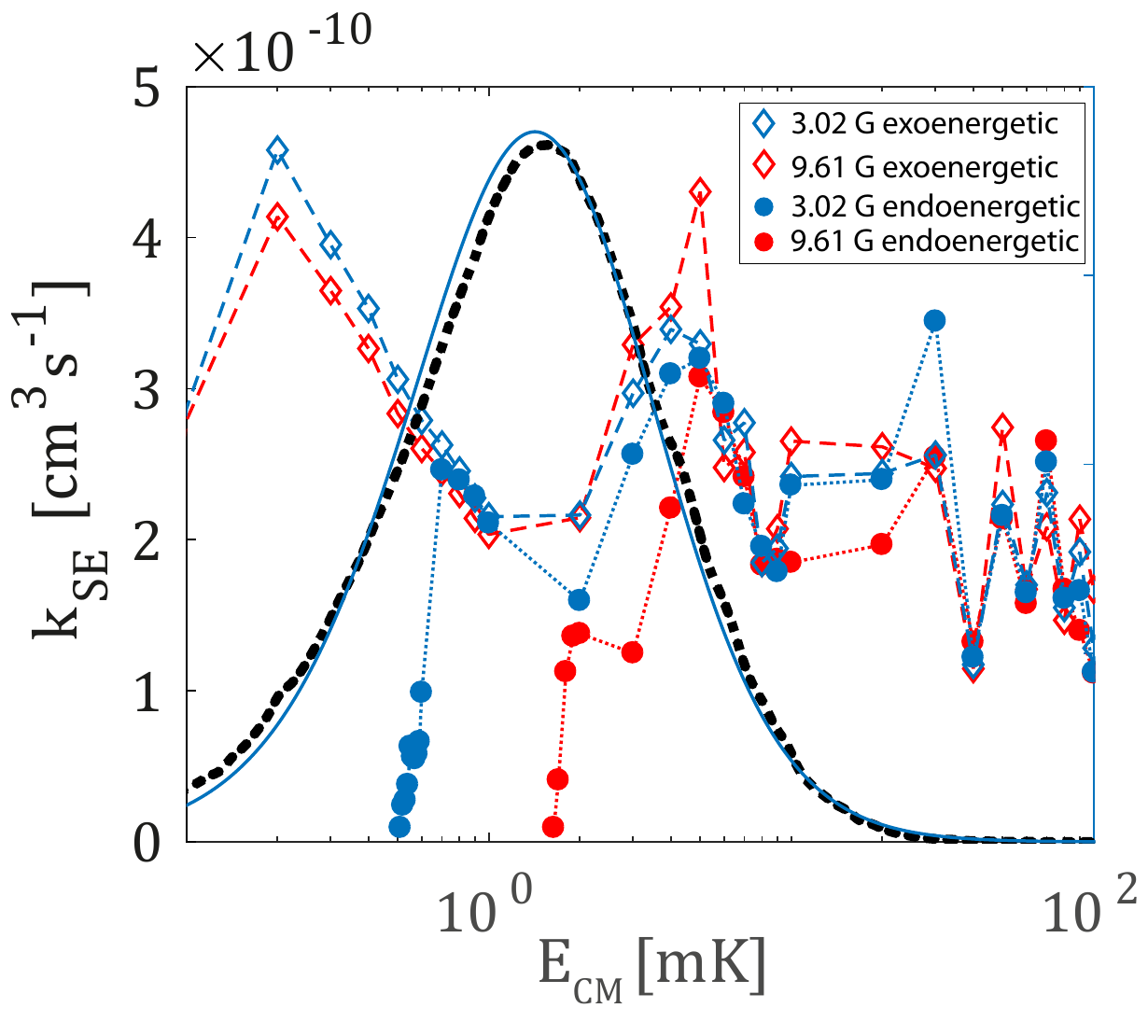}
	\caption{\label{Supplementary figure 3.}Supplementary figure 3. Energy distribution in the laboratory frame extracted from the simulation (black dashed line). Tsallis fit $n=4.04$ $T=\SI{0.43}{mK}$ (blue line). Endoenergetic $\ket{1,0}_\mathrm{Rb}\otimes\ket{\downarrow}_{\mathrm{Sr}^+}\rightarrow\ket{1,-1}_\mathrm{Rb}\otimes\ket{\uparrow}_{\mathrm{Sr}^+}$ (circles) and exoenergetic  $\ket{1,0}_\mathrm{Rb}\otimes\ket{\uparrow}_{\mathrm{Sr}^+}\rightarrow\ket{1,1}_\mathrm{Rb}\otimes\ket{\downarrow}_{\mathrm{Sr}^+}$ (diamonds) SE rate for $B=\SI{3.02}{\gauss}$ (blue) and $B=\SI{9.61}{\gauss}$ (red).}
\end{figure}
\bibliography{cite.bib}
\end{document}